\documentclass[twocolumn,showpacs,preprintnumbers,amsmath,amssymb]{revtex4}

\usepackage{graphicx}
\usepackage{dcolumn}
\usepackage{bm}

\begin{document}

\preprint{}

\title{Effect of photoions on the line shapes of the F\"orster resonance and microwave transitions in cold rubidium Rydberg atoms}
\author{D.~B.~Tretyakov}
\author{I.~I.~Beterov}
\author{V.~M.~Entin}
\author{E.~A.~Yakshina}
\author{I.~I.~Ryabtsev}
  \email{ryabtsev@isp.nsc.ru}
\affiliation{Institute of Semiconductor Physics SB RAS, Prospekt Lavrentyeva 13, 630090 Novosibirsk, Russia }
\author{S.~F.~Dyubko}
\author{E.~A.~Alekseev}
\author{N.~L.~Pogrebnyak}
\affiliation{Radioastronomy Institute NAS Ukraine, Ul. Krasnoznamyonnaya 4, 61002 Kharkov, Ukraine}
\author{N.~N.~Bezuglov}
\affiliation{Saint-Petersburg State University, Ul. Ulyanovskaya 3, Peterhof, 198904 S.-Petersburg, Russia}
\author{E.~Arimondo}
\affiliation{Universita di Pisa, Largo Pontecorvo 3, I-56127 Pisa, Italy}

\date{May 25, 2011}

\begin{abstract}
Experiments on the spectroscopy of the F\"orster resonance Rb(37\textit{P})+Rb(37\textit{P})$\rightarrow$ Rb(37\textit{S})+Rb(38\textit{S}) and microwave transitions $nP\to n'S,\; n'D$ between Rydberg states of cold Rb atoms in a magneto-optical trap have been performed. Under ordinary conditions, all spectra exhibited a 2$-$3 MHz line width independently of the interaction time of atoms with each other or with microwave radiation, although the ultimate resonance width should be defined by the inverse interaction time. Analysis of the experimental conditions has shown that the main source of the line broadening was the inhomogeneous electric field of cold photoions appeared at the excitation of initial Rydberg \textit{nP} states by broadband pulsed laser radiation. Using an additional pulse of the electric field, which rapidly removed the photoions after the laser pulse, lead to a substantial narrowing of the microwave and F\"orster resonances. An analysis of various sources of the line broadening in cold Rydberg atoms has been conducted.
\end{abstract}

 \maketitle

\section{Introduction}

Cold atoms in highly excited Rydberg states with a principal quantum number $n\gg1$ are of interest for basic research and practical applications [1]. In particular, spectroscopy of cold atoms opens up new possibilities for improving the measurement accuracy of the energy and spectroscopy parameters of atoms due to eliminating the Doppler effect [2-4], while the cold Rydberg atoms themselves can be used for implementing logic gates of a quantum computer [5,6].

In this kind of research, an important role is played by the line width of optical and microwave transitions between Rydberg states, because this line width determines the spectral resolution in spectroscopic measurements and the coherence time when performing quantum operations. In the absence of external electromagnetic fields, the ultimate line width is equal to the inverse lifetime of Rydberg states [7,8] and amounts to 1$-$10 kHz for $n>30$. However, this line width is hard to observe experimentally, because Rydberg states have a hyperfine structure (hundreds of kilohertz for \textit{nS} states and tens of kilohertz for \textit{nP} states at $n\sim 40$ [3,4]). The ultimate resolution is also restricted by the finite interaction time $t_{0} $ with resonance radiation, which gives a Fourier width of resonances $\sim 1/t_{0} $. In addition, experimental investigations have revealed a number of problems associated with the broadening of spectral lines in external electromagnetic fields.

First, in magneto-optical traps (MOTs), an inhomogeneous quadrupole magnetic field is used for trapping and cooling atoms. The typical values of the field gradient are 10$-$15 G/cm; therefore, the variation of the magnetic field over a cloud of cold atoms of $\sim1$ mm in size is about 1 G. This field acts on Rydberg atoms and splits their levels by 1$-$3 MHz due to the Zeeman effect. To reduce this effect on line broadening, one applies either a short-term switching off of the magnetic field during measurements [3,4], or a localization of a small excitation volume of Rydberg atoms near the center of the cloud of cold atoms, where the field vanishes [9]. This allows one to reduce the contribution of the inhomogeneity of the MOT magnetic field to the resonance line width to 10-100 kHz. It was also pointed out that, when studying two-photon microwave transitions between Rydberg states with identical magnetic structures, it is possible to observe narrow resonances even without switching off the magnetic field, because the energy levels of these transitions are identically shifted by the magnetic field [2].

\begin{figure*}
\includegraphics[scale=0.8]{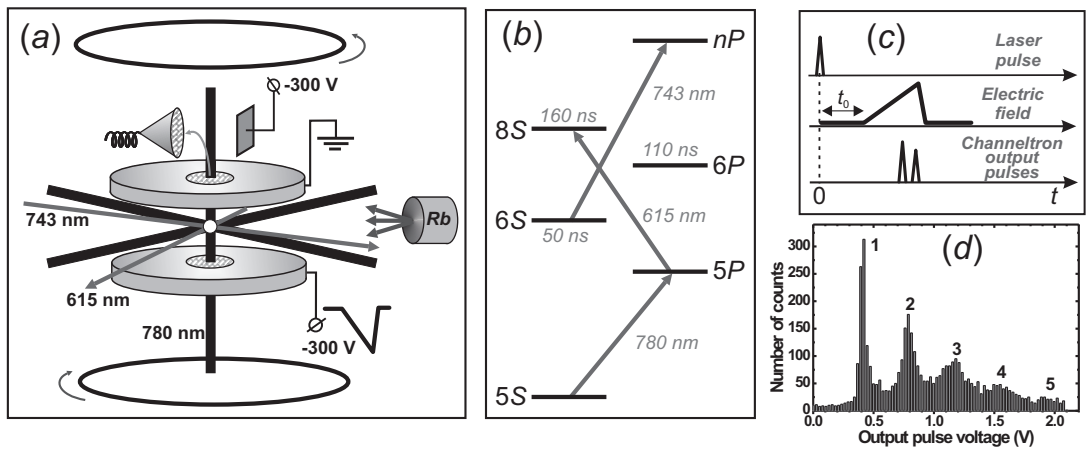}
\caption{\label{Fig1} (\textit{a}) Scheme of the experiment with cold Rb Rydberg atoms in a magneto-optical trap. (\textit{b}) Scheme of the three-step laser excitation of Rydberg \textit{nP} states in Rb atoms. (\textit{c}) Timing diagram of the pulses in experiments with selective field ionization (SFI) of Rydberg atoms. (\textit{d}) Histogram of the amplified output pulses of the channeltron VEU-6. The observed peaks correspond to 1$-$5 electrons detected from Rydberg atoms by SFI.}
\end{figure*}

Second, transition lines in cold Rydberg atoms are broadened by spurious electromagnetic fields, which always exist in the experiments. Rydberg atoms exhibit the highest sensitivity to spurious electric fields, because the polarizability of Rydberg levels grows as $n^{7} $ [1]. Electric fields of a few mV/cm may be enough to shift and split the transition levels by several or even tens of megahertz. If a spurious field is spatially homogeneous, it can be compensated for with the use of a system of additional electrodes by minimizing the Stark shifts and line broadening [3,4]. If the field is inhomogeneous, then it can be compensated for only partially even in a small interaction volume; therefore, the transition lines experience residual broadening.

Third, experiments with cold Rydberg atoms in a MOT revealed a spontaneous evolution of these atoms into an ultracold plasma upon reaching a certain critical density, which depends on \textit{n} [10]. For the density of atoms of higher than $10^9$ cm$^{-3}$, plasma is formed within a few microseconds in spite of the low kinetic energy of the atoms. Primary charged particles can be produced either due to the photoionization of Rydberg atoms by laser and background thermal radiation, or as a result of collisions of cold Rydberg atoms in the case of attracting potential between them [11]. From the viewpoint of precision spectroscopy, it is important that, in the absence of an extracting electric field, charged particles of the ultracold plasma stay near the Rydberg atoms for a long time. The inhomogeneous electric field of these particles may lead to the broadening and shift of spectral lines in the Rydberg atoms due to the Stark effect [12].

In the present study, we investigated the effect of cold photoions on the spectra of the F\"orster energy-exchange resonance Rb(37\textit{P})+Rb(37\textit{P})$\rightarrow$Rb(37\textit{S})+Rb(38\textit{S}) and microwave transitions $nP\to n'S,\; n'D$ between Rydberg states of cold rubidium atoms in a MOT. In our experiments, we have revealed that one of the main sources of line broadening is the inhomogeneous electric field of cold photoions that are generated under the excitation of Rb(\textit{nP}) Rydberg atoms by broadband pulsed laser radiation. By using the spectroscopy of microwave transitions and exchange interactions, we implemented a nondestructive spectroscopic method for the diagnostics of an ultracold plasma in a gas of cold rubidium Rydberg atoms in a MOT. The applicability of this method was first demonstrated in [12]. In contrast to [12], in the present experiments we obtain higher sensitivity to weak electric fields from a small amount of photoions and higher spatial resolution owing to the localization of the excitation volume in the geometry of tightly focused crossed laser beams.

\section{EXPERIMENTAL SETUP}

A detailed description of the experimental setup was given in our paper [9]. The experiments are carried out with cold Rb atoms trapped in a MOT, which is shown in Fig.~1(a). The trap consists of a vacuum chamber with optical windows, a heated source of Rb atoms, anti-Helmholtz coils for creating a three-dimensional gradient of a magnetic field of 10$-$15 G/cm at the center of the MOT, and a laser cooling system based on two semiconductor lasers with external cavities tuned to a wavelength of 780 nm. The atoms are cooled by three orthogonal pairs of light waves. The cooling laser is tuned with a red detuning of $\sim$20 MHz to the closed transition $5S_{1/2} (F=3)\to 5P_{3/2} (F=4)$ of the $^{85} {\rm Rb}$ isotope, and the repumping laser to the transition $5S_{1/2} (F=2)\to 5P_{3/2} (F=3)$. After the fine tuning of the wavelengths of the lasers, a cloud of cold atoms with a size of 0.5$-$1 mm and temperature of 100$-$300 $\mu $K arises at the center of the trap. In our experiments, about $10^6$ atoms of $^{85} {\rm Rb}$ are trapped in the MOT, which corresponds to a number density of $10^9$ cm$^{-3}$.

The excitation and detection of Rydberg atoms occur in the space between two stainless-steel plates with holes of 10 mm in diameter at the centers [Fig.~1(a)]. To form a homogeneous electric field, the holes are closed by optically transparent (85\% transparency) metal meshes. The distance between the plates is 10 mm. The electric field is used for the spectroscopy of the Stark effect and for detecting Rydberg atoms by the selective field ionization (SFI) method [1]. The electrons generated upon ionization are accelerated by the electric field, fly through the upper mesh, and are directed by a deflecting electrode into the input horn of a channel electron multiplier VEU-6. Pulse signals from the output of this multiplier are processed by a fast analog-to-digital converter, a box-car integrator, and a computer. This allows one to monitor the number of atoms and the populations of the Rydberg states in a wide range of the principal quantum number \textit{n}.

The excitation of cold Rb atoms to the Rydberg states \textit{nP} (\textit{n}=30-60) is performed by a three-step scheme [Fig.1(b)]. The first step $5S_{1/2} \to 5P_{3/2} $ is excited by a cooling laser operating in a continuous-wave mode. At the second step $5P_{3/2} \to 8S_{1/2} $ the radiation of a pulsed Rhodamine 6G dye laser (with a wavelength of 615 nm) is used at a pulse repetition rate of 5 kHz. The 8\textit{S} state has a lifetime of 160 ns and rapidly decays into lower lying \textit{P} states, including the 6\textit{P} state. In turn, this state has a lifetime of 110 ns and rapidly populates the state 6\textit{S}\,, which has a lifetime of 50 ns. At the third step the Rydberg \textit{nP} states are excited from the 6\textit{S} state by the radiation of a pulsed titanium-sapphire laser with a wavelength of 743 nm. The width of both laser pulses is about 50 ns and the lasers themselves are time synchronized. The radiation of the second- and third-step lasers is focused onto a cloud of cold atoms in the crossed-beam geometry [Fig.~1(a)] with waist diameters of 9$-$10 $\mu $m. In the region of intersection of the focused beams, an effective excitation volume of Rydberg atoms with a size of 20$-$30 $\mu $m is formed, which depends on the relative position of the waists and on whether or not the transitions are saturated.

The diagnostics of cold Rb Rydberg atoms in a MOT is carried out by the method of microwave spectroscopy [1]. A backward-wave-tube oscillator G4-142 is used as a source of microwave radiation. The frequency of the oscillator is locked to a quartz frequency synthesizer and is swept within 53$-$80 GHz range at a line width of 20$-$50~kHz. The radiation is fed through the MOT window. The spectra of microwave transitions give information on the presence of external electric and magnetic fields and on their spatial distribution, because the excitation volume can be shifted within the cold atoms cloud.

The time diagram of signals in the detection system is shown in Fig.~1(c). After every laser pulse that excites a part of cold atoms to the initial \textit{nP} Rydberg state, atoms freely interact with each other or with microwave radiation during a period of $t_0=$1$-$10 $\mu $s. Then a ramp of the ionizing electric field with a rise time of about 2~$\mu $s is switched on. Depending on the state of a Rydberg atom, the ionization occurs at different instants of time after a laser pulse. Then a pulsed ionization signal is detected at the output of VEU-6 by means of two gates that correspond to the initial \textit{nP} and final $n'L$ states of a Rydberg atom. The number of electrons detected from a single laser pulse is determined by the number of Rydberg atoms in the excitation region and by the detection efficiency of VEU-6 [13]. Figure 1(d) shows a histogram of the amplitudes of output signals of VEU-6. The histogram shows several peaks that correspond to different numbers (from 1 to 5) of detected Rydberg atoms. After every laser pulse the data acquisition system measures the amplitude of the output signal of the VEU-6 in both detection channels (for the initial and final states); then, by a previously measured histogram, the system determines the number of atoms detected in each channel, and upon accumulating data from $10^3-10^5$ laser pulses sorts the signals according to the number of atoms and calculates the probability of transition from the initial to the final Rydberg state.

\section{SPECTROSCOPY OF A F\"ORSTER RESONANCE IN THE PRESENCE OF COLD PHOTOIONS}

A F\"orster resonance, or the resonance energy-exchange transfer, occurs due to the dipole-dipole interaction between neighboring Rydberg atoms and is observed in the case of two or more atoms when the energy intervals of the up and down transitions from the initial state are equal [1]. In our experiments, we used the 37$P_{3/2}$ state as the initial state; for this state, the resonance interaction of two atoms occurs according to the scheme [9]

\noindent 

\begin{equation} \label{Eq1} 
\begin{array}{l} {{\rm Rb}(37P_{3/2} )+{\rm Rb}(37P_{3/2} )\to } \\ \\{{\rm Rb}(37S_{1/2} )+{\rm Rb}(38S_{1/2} ).} \end{array}
\end{equation} 

\noindent The exact energy resonance for this process is achieved by the Stark tuning of Rydberg levels in the electric field [Fig.~2(a)]. As a result of interaction, one of the atoms goes to the lower state 37$S_{1/2}$ and the other atom simultaneously goes to the higher state 38$S_{1/2}$. The interaction may involve not only two but a larger number of atoms located in the excitation volume, provided that they are situated sufficiently closely.

\begin{figure*}
\includegraphics[scale=0.8]{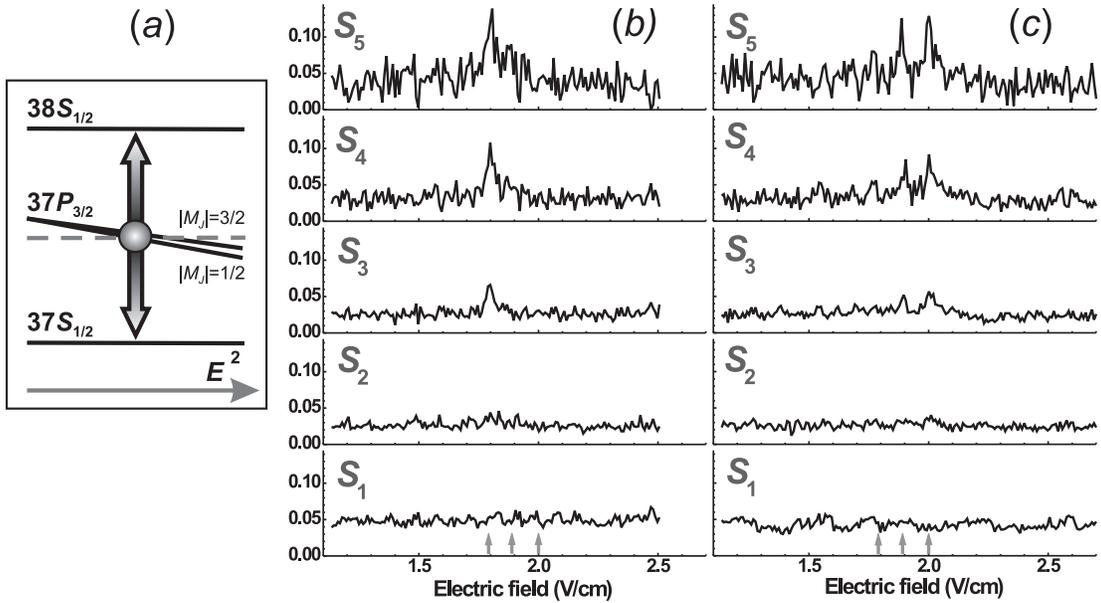}
\caption{\label{Fig2} (\textit{a}) Scheme of tuning the double Stark resonance $37S_{1/2} -37P_{3/2} -38S_{1/2} $ in Rb atoms in the electric field \textit{E}=1.8-2.0~V/cm. (\textit{b}) Records of the spectra of the F\"orster resonance Rb(37\textit{P})+Rb(37\textit{P})$\rightarrow$Rb(37\textit{S})+Rb(38\textit{S}) at $\pi $-polarization of the exciting laser and selective detection of 1$-$5 Rydberg atoms. The vertical arrows indicate the calculated positions of the three resonances. (\textit{c}) The same as in (\textit{b}) for $\sigma $-polarization of the exciting laser.}
\end{figure*}

To calculate the evolution of the populations of Rydberg states, we solve a quantum-mechanical problem for a quasimolecule formed by \textit{N} interacting Rydberg atoms [14,15]. In [15] we have shown that for a weak dipole-dipole interaction of frozen Rydberg atoms the evolution of the population of the final state 37\textit{S} in each atom and the line shape of the F\"orster resonance are described by the expression

\noindent 

\begin{equation} \label{Eq2} 
\rho _{N} (t_{0} )\approx \frac{1}{N} \, \; \frac{V_{N}^{2} }{V_{N}^{2} +\Delta ^{2} /4} \sin ^{2} \left(t_{0} \; \sqrt{V_{N}^{2} +\Delta ^{2} /4} \right), 
\end{equation} 

\noindent where $V_{N} $ is the total energy of the dipole-dipole interaction of all pairs of atoms (in the frequency scale), $\Delta =\left(2E_{37P} -E_{37S} -E_{38S} \right)/\hbar $ is the electric-field controlled detuning from the exact energy resonance, $t_{0} $ is the interaction time, and $1/N$ is the normalization constant. This formula is similar to the formula describing the Rabi oscillations in a two-level atom; as $V_{N} \to 0$, the line shape is given by the Fourier spectrum of a square pulse of width $1/t_{0} $. Accurate numerical calculations have shown that Eq.\eqref{Eq2} gives correct results for $\rho _{N} (t_{0} )<0.1$.

The change in the states of atoms is detected by the SFI method by measuring the population of the final state 37$S_{1/2}$. The distinctive feature of our experiments is the possibility of determining the number \textit{N} of Rydberg atoms and their states after every laser pulse. As a result, we measure the following signals:

\begin{equation} \label{Eq3} 
S_{N} =\frac{n_{N} (37S)}{n_{N} (37P)+n_{N} (37S)+n_{N} (38S)}\; . 
\end{equation} 

\noindent Here $n_{N} (nL)$ is the total number of Rydberg atoms in state \textit{nL} that are detected during the measurement time in the case of \textit{N} Rydberg atoms. In fact, the signal $S_N$ represents the mean probability of transition in each atom after the interaction with \textit{N}$-$1 surrounding atoms; i.e., $S_{N} =\rho _{N} $ if an ideal SFI detector with a detection efficiency $\eta =1$ is used. The efficiency of our detection system is 65\%; therefore, actually $S_N$ represents a mixture of signals $\rho _{i} $ from a larger number of actually excited atoms $i\ge N$ [14].

\begin{figure*}
\includegraphics[scale=0.6]{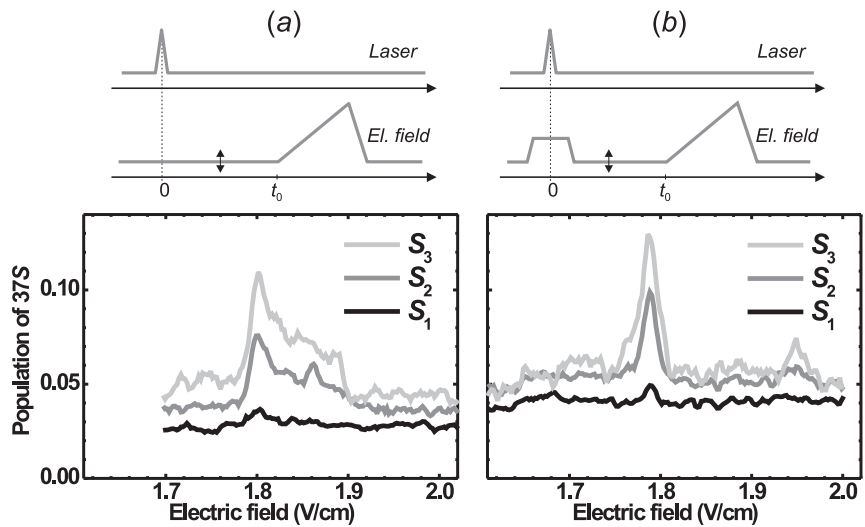}
\caption{\label{Fig3} Experimental records of the spectrum of the F\"orster resonance Rb(37\textit{P})+Rb(37\textit{P})$\rightarrow$Rb(37\textit{S})+Rb(38\textit{S}) in cold Rb atoms for 1$-$3 detected Rydberg atoms at $\pi $-polarization of the exciting laser radiation: (\textit{a}) In the presence of cold photoions, which appeared at laser excitation of the 37\textit{P} Rydberg state. The electric field of the photoions broadens the F\"orster resonance. (\textit{b}) Additional electric pulse (5 V/cm, 2~$\mu $s) at the moment of laser excitation rapidly extracts photoions. A narrow F\"orster resonance in the field 1.79 V/cm is observed.}
\end{figure*}

Since the state $37P_{3/2} $ is split in the electric field into two sublevels, $|M_{J} |=1/2$ and $|M_{J} |=3/2$, where $M_J$ is a moment projection onto the direction of the electric field, in general three F\"orster resonances should be observed for the electric fields of 1.79, 1.89, and 2.0 V/cm (these values are obtained from the numerically calculated Stark diagram of Rydberg levels). In our first paper [9] we could not resolve these resonances due to the broadening and overlapping of individual resonances due to the inhomogeneity of the electric field and spurious electromagnetic fields, as well as due to the weakness of interatomic interaction in the excitation volume with a size of $\sim$100 $\mu $m.

In the subsequent experiments, the excitation volume was first reduced by making a small separation of the waists of two exciting laser beams, so that they only slightly overlapped, forming a reduced effective excitation volume with a size of 30$-$40 $\mu $m. In spite of the fact that such a technique reduces the average number of atoms in the excitation volume and degrades the signal-to-noise ratio, we could record individual components of the F\"orster resonance. These records are shown in Figs.~2(b) and 2(c) for $\pi $ and $\sigma $ polarizations of the exciting laser radiation. In the case of $\pi $ polarization, only the Stark sublevel $37P_{3/2} (|M_{J} |=1/2)$ is excited from the $6S_{1/2} $ state; therefore, only a single F\"orster resonance is observed in a field of 1.79 V/cm. In the case of $\sigma $ polarization, both Stark sublevels $|M_{J} |=1/2$ and $|M_{J} |=3/2$ are excited, and the population of the latter sublevel turns out to be three times higher; therefore, the F\"orster resonance in a field of 1.79 V/cm is three times weaker, and one can clearly observe only resonances in the fields of 1.89 and 2.9 V/cm over the background of noise. In this experiment, the width of each resonance was 30$-$40 mV/cm, which corresponds to 3$-$4 MHz when recalculated to the frequency scale. Thus, the spectral resolution was still several times poorer than that calculated by Eq.~\eqref{Eq2}, which should be about 1 MHz with regard to the unresolved hyperfine structure of Rydberg levels (about 0.7 MHz in total) and the finite interaction time (0.3 MHz for an interaction time of 3 $\mu $s). This means that other sources of line broadening listed in the Introduction contribute to the width of the resonances.

To date, we have undertaken a number of efforts that have allowed us to reduce the excitation volume to 20$-$30~$\mu $m by tighter focusing the laser beams and more carefully superimposing their waists. We have also undertaken special measures to eliminate spurious electric fields associated with ground loops and stray fields in the detection system. This has allowed us to substantially reduce the effect of inhomogeneity of the electric field and improve the spectral resolution when detecting F\"orster resonances [14].

To analyze additional sources of line broadening, we recorded a single F\"orster resonance (at $\pi $ polarization of laser radiation) with a smaller step in the electric field scale and with a larger accumulation time to improve the signal-to-noise ratio (Fig.~3). The F\"orster resonance was observed in a weak dc electric field, which was varied near the value of 1.79 V/cm. The interaction time $t_{0} =3$ $\mu $s was defined by the time interval between a laser pulse and the instant when a strong ionizing field for SFI is switched on. Some of the records showed a strongly asymmetric broadening of the F\"orster resonance for higher values of the electric field [Fig.~3(a)]. While the resonance amplitude strongly depended on the number of detected Rydberg atoms, its width and shape were constant. This indicates that the asymmetric broadening is attributed to an inhomogeneous electric field of unknown nature, rather than to the interaction between Rydberg atoms.

According to our previous experiments [9], the measured inhomogeneity of the electric field formed by the plates of the detection system in the excitation volume of 100 $\mu $m in size was not greater than 0.5\%, which amounts to 10 mV/cm for a field of 2 V/cm. Since in our present experiment we have even smaller excitation volume of 20$-$30~$\mu $m in size, the field inhomogeneity cannot be greater than 2$-$3~mV/cm. However, the broadening of resonances shown in Fig.~3(a) amounts to 50 mV/cm. The only source of such an inhomogeneous electric field can be the charged particles situated directly in the excitation volume.

It was suggested that the source of broadening of the F\"orster resonance is the inhomogeneous electric field of photoions generated under a pulsed excitation of Rb Rydberg atoms by a broadband laser radiation via the three-step scheme shown in Fig.~1(b). It was observed in the experiments that laser pulses used on the second and third steps produced a strong photoionization signal during a laser pulse if one additionally applies an extracting electric field of higher than 10 V/cm. Since photoionization does not require resonance radiation, all photons of laser pulses take part in the photoionization, whereas only resonance photons (the absorption line width $\sim$5 MHz) from the entire spectrum of the pulsed lasers (the line width of the lasers $\sim$10 GHz) take part in the excitation of Rydberg states. The photoionization signal turns out to be comparable with, or even greater than, the signals from the Rydberg atoms. The wavelengths of the lasers of the second and third steps allowed one to photoionize all the states above the  6\textit{S} state that are populated under a spontaneous decay of the state 8\textit{S}\,. Therefore, it seems impossible to calculate the total photoionization probability because of the unknown distribution of populations over these levels and their time dynamics.

Photoionization gives rise to the appearance of free electrons and cold photoions Rb$^+$. A weak electric field of about 1.79 V/cm applied to observe the F\"orster resonance extracts photoelectrons from the excitation volume of 20 $\mu $m in a time of about 1 ns, whereas the extraction of cold photoions takes about 0.5 $\mu $s. This time period is comparable with the interaction time $t_{0} =3$ $\mu $s of atoms with each other. Due to the Stark effect, the cold ions lead to the deviation of the frequencies of atomic transitions during the interaction time. The asymmetric broadening of the F\"orster resonance in Fig.~3(a) is about 50 mV/cm and it immediately yields the mean electric field of photoions, which start to move to the negatively charged plate of the detection system under our external field. Thus, the F\"orster resonance spectroscopy allows one to measure the mean field of an ultracold plasma in a gas of cold Rydberg atoms.

Our method for detecting charged particles allows one to measure also the number (from 0 to 10) of photoions produced after each laser pulse. The probability of generation of a certain number of ions is described by the Poisson statistics. Figure 3(a) shows that the left wing of the F\"orster resonance is not broadened; this may imply that the signal of this wing corresponds to the cases when photoions are not generated. The asymmetric signal on the right wing corresponds to the cases when photoions are first generated and then are slowly moved by the control electric field to one side. The possibility of the direct measurement of the mean electric field and of the number of photons is of interest for studying the initial stage of formation of ultracold plasma.

To check the assumption on the presence of photoions, we suggested that during a laser pulse one should apply an additional electric-field pulse of amplitude 5 V/cm and duration of 2 $\mu $s; this pulse should extract photoions in a time of 0.2$-$0.3 $\mu $s [Fig.~3(b)]. During this pulse, atoms are far-detuned from the F\"orster resonance and do not interact with each other, while photoions are completely removed from the excitation volume. After the end of the pulse, the electric field decreases to the resonance value, and then atoms interact with each other during $t_{0} =3$ $\mu $s. The experimental record of the F\"orster resonance demonstrated in Fig.~3(b) shows that photoions indeed cease to affect the line shape and that the resonance becomes nearly symmetric (the residual asymmetry is associated with the exponential transient process during switching off the extracting pulse). For a one-atom signal $S_{1} $ the resonance width decreases to $16.4\pm 0.3$ mV/cm, which corresponds to $1.94\pm 0.04$ MHz on the frequency scale. For multiatom signals the resonance width turns out to be larger due to the larger energy of interatomic interactions. A detailed analysis of the shape of the F\"orster resonance as a function of the number of detected Rydberg atoms was carried out in [14,15]; therefore, here we do not discuss this issue.

\section{SPECTROSCOPY OF MICROWAVE TRANSITIONS IN THE PRESENCE OF COLD PHOTOIONS}

To absolutely calibrate the electric field strength in experiments on spectroscopy of the F\"orster resonance ${\rm Rb}(37P_{3/2} )+{\rm Rb}(37P_{3/2} )\to {\rm Rb}(37S_{1/2} )+{\rm Rb}(38S_{1/2} )$, we used a resonant microwave radiation. The exact calibration was needed to measure and compensate for the spurious electric fields in the detection system. As can be seen from Fig.~2(a), for the calibration the frequency of the microwave field should be tuned to the frequency of the exact double resonance $37S_{1/2} -37P_{3/2} -38S_{1/2} $ that arises in an electric field of 1.79$-$2.0 V/cm. When the electric field strength is varied, the measured signal of the population of the state $37S_{1/2} $ should exhibit a resonance similar to the F\"orster resonance in Fig.~3; however, the amplitude of this resonance is determined by the intensity of the microwave field rather than by the energy of interatomic interactions. To prevent additional broadening of the resonance due to these interactions, one should consider a one-atom signal $S_{1} $\,.

Figure 4 shows the records of such resonances for a microwave field frequency of 81.072 GHz and for $\pi $ polarization of the exciting laser radiation. In the presence of photoions [Fig.~4(a)], the resonance peak at 1.79~V/cm is broadened asymmetrically. Moreover, a forbidden component arises in the spectrum in a field of 2.0 V/cm, that can only be observed if the Stark sublevel $37P_{3/2} (|M_{J} |)=3/2$ is populated. However, in the case of $\pi $ polarization of the laser radiation this sublevel should not be populated; therefore, one should conclude that the electric field of photoions in this experiment had a transverse component with respect to the dc electric field in the detection system. The transverse field lead to the precession of the magnetic moment and ultimately lead to the population of the state $37P_{3/2} (|M_{J} |)=3/2$, as we pointed out in [16]. This fact is confirmed by the record of a microwave resonance in the absence of photoions [Fig.~4(b)]. In this case, one observes a single narrow peak in a field of 1.79 V/cm, which was used for the absolute calibration of the electric field strength in the detection system in our experiments.

\begin{figure*}
\includegraphics[scale=0.6]{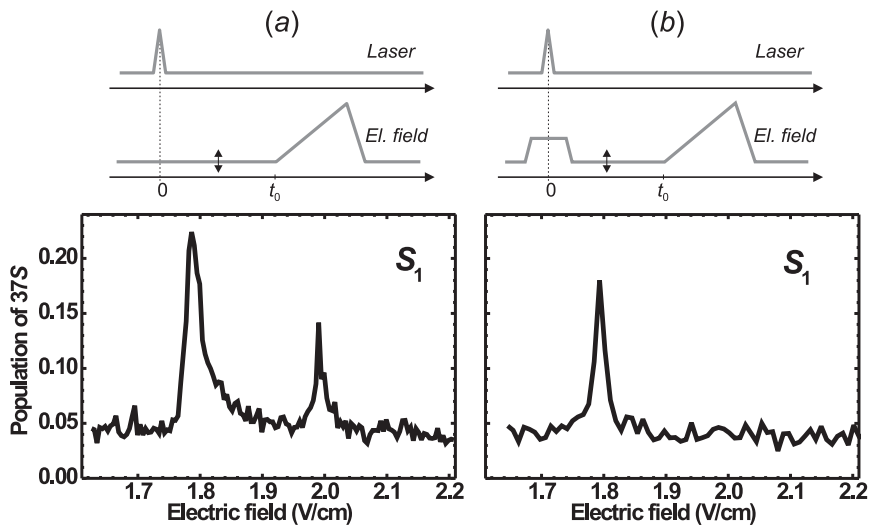}
\caption{\label{Fig4} Calibration of the position of the F\"orster resonance in a microwave field with the 81.072 GHz frequency at $\pi $-polarization of the exciting laser radiation: (\textit{a}) In the presence of cold photoions, which appeared at laser excitation of the 37\textit{P} Rydberg state. The electric field of the photoions broadens the microwave resonance and causes an appearance of the forbidden component in the field 2.0 V/cm. (\textit{b}) Additional electric pulse (5 V/cm, 2~$\mu $s) at the moment of laser excitation rapidly extracts photoions. A single narrow microwave resonance in the field 1.79 V/cm is observed.}
\end{figure*}

We also carried out experiments on the spectroscopy of one-photon microwave transitions 37\textit{P}$-$37\textit{S}\,, 40\textit{P}$-$40\textit{S}\,, 45\textit{P}$-$43\textit{D}\,, 50\textit{P}$-$50\textit{D}\,, 55\textit{P}$-$55\textit{D}\,, and 60\textit{P}$-$61\textit{D} between Rydberg states of cold Rb atoms in a MOT. The choice of these transitions was motivated by the fact that their frequencies fall into the tuning range (50$-$80 GHz) of the G-142 microwave generator. The aim of the experiments was to study the possibility of excitation and diagnostics of high Rydberg states for further investigations in the field of quantum information, as well as to improve the values of quantum defects and transition frequencies in Rb Rydberg atoms. To this end, we carried out a detailed analysis of the line shapes of microwave resonances.

As already discussed in the Introduction, the ultimate spectral resolution in experiments on the spectroscopy of cold Rydberg atoms is determined by their finite lifetime or finite interaction time between atoms and microwave radiation. Since the interaction time is usually limited to the values of 1$-$10 $\mu $s and the lifetimes are greater than tens of microseconds for $n>30$ [7,8], the ultimate line width is mainly determined by the Fourier width of microwave pulses and amounts to 0.1$-$1 MHz. The probability of a one-photon transition and its line shape for a square microwave pulse of duration $t_{0} $ are described by the well-known formula

\begin{equation} \label{Eq4} 
\rho (t_{0} )\approx \frac{\Omega ^{2} }{\Omega ^{2} +\delta ^{2} } \sin ^{2} \left(\frac{t_{0} }{2} \sqrt{\Omega ^{2} +\delta ^{2} } \right), 
\end{equation}

\noindent where $\Omega $ is the Rabi frequency and $\delta $ is the frequency detuning from the exact resonance. This formula describes Rabi oscillations in a two-level atom, and for $\Omega \to 0$ the line shape is given by the Fourier spectrum of a square pulse of width $1/t_{0} $. In our experiments, the interaction time was chosen equal to $t_{0} =3$ $\mu $s, which corresponds to a Fourier width of 0.33 MHz. However, in the first experiments, we observed approximately equal line widths for all microwave transitions, which was 2$-$3 MHz irrespective of the interaction time. Moreover, the measured transition frequencies were appreciably shifted with respect to the calculated values obtained with the use of quantum defects from [2].

As the main mechanism of the shift and broadening of microwave transition lines, we considered the Stark effect in spurious electric fields. To study this phenomenon, we applied an additional small dc voltage of either polarity to the plates of the detection system that formed a homogeneous electric field in the vertical direction; this small dc field was used to compensate for the dc component of the spurious electric field. We found an unaccounted electric field of strength 100$-$150 mV/cm in the detection system, which penetrated into the excitation region from the deflecting electrode through the metal mesh of the detection system [see Fig.~1(a)]. The compensation of the vertical component of this field allowed us to significantly reduce the shifts and broadenings of microwave resonances to values of 2 MHz. As an example, Fig.~5 shows the spectra of the microwave resonance $37P_{3/2} \to 37S_{1/2} $ for various values of the compensating dc voltage \textit{U}. The figure also presents the resulting electric field calculated by the shift of the resonance with respect to the calculated position. Note that even for the minimum shift of the microwave resonance Fig.~5 shows broadening due to the inhomogeneous magnetic field of the MOT. Later, this broadening was eliminated by adjusting the position of the excitation volume to the point of zero magnetic field [9]. 

\begin{figure}
\includegraphics[scale=0.7]{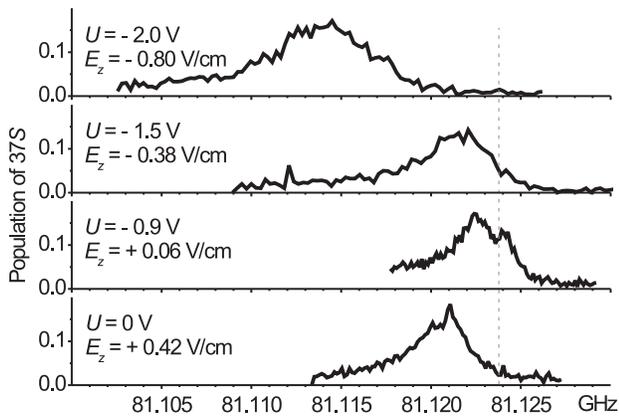}
\caption{\label{Fig5} Spectrum of the microwave resonance $37P_{3/2} \to 37S_{1/2} $ between Rydberg states of Rb atoms at various values of the compensating voltage \textit{U} in the detection system. Also shown are the values of the resulting vertical component of the electric field calculated from the shift of the resonance with respect to the theoretical position (dashed line).}
\end{figure}

Moreover, often additional spurious electric fields appeared to the end of the experiments; these fields were presumably attributed to the deposition of Rb atoms on the metal parts of the detection system and to the rise of a contact potential difference. The elimination of these fields required a periodic adjustment of the compensating voltage in the detection system. This phenomenon was earlier observed in a number of other publications on microwave spectroscopy of Rydberg atoms [17].

Figure 6(a) shows the spectra of the microwave transitions $37P_{3/2} \to 37S_{1/2} $, $45P_{3/2} \to 43D_{3/2} $, and $50P_{3/2} \to 50D_{3/2} $ obtained after the adjustment of the position of the excitation volume to a zero magnetic field point and after the compensation of the vertical component of the spurious electric field, but without the extracting pulse for photoions. The arrows indicate the calculated positions of the resonances according to the quantum defects from [2]. One can see that the measured transition frequencies differ from the calculated ones by at most 1~MHz. Nevertheless, the observed resonance widths are still rather large and amount to 1.5$-$2 MHz, which provides evidence for the existence of a time-dependent electric field of photoions generated under laser excitation, just as in the case of the spectroscopy of the F\"orster resonance.

\begin{figure*}
\includegraphics[scale=0.6]{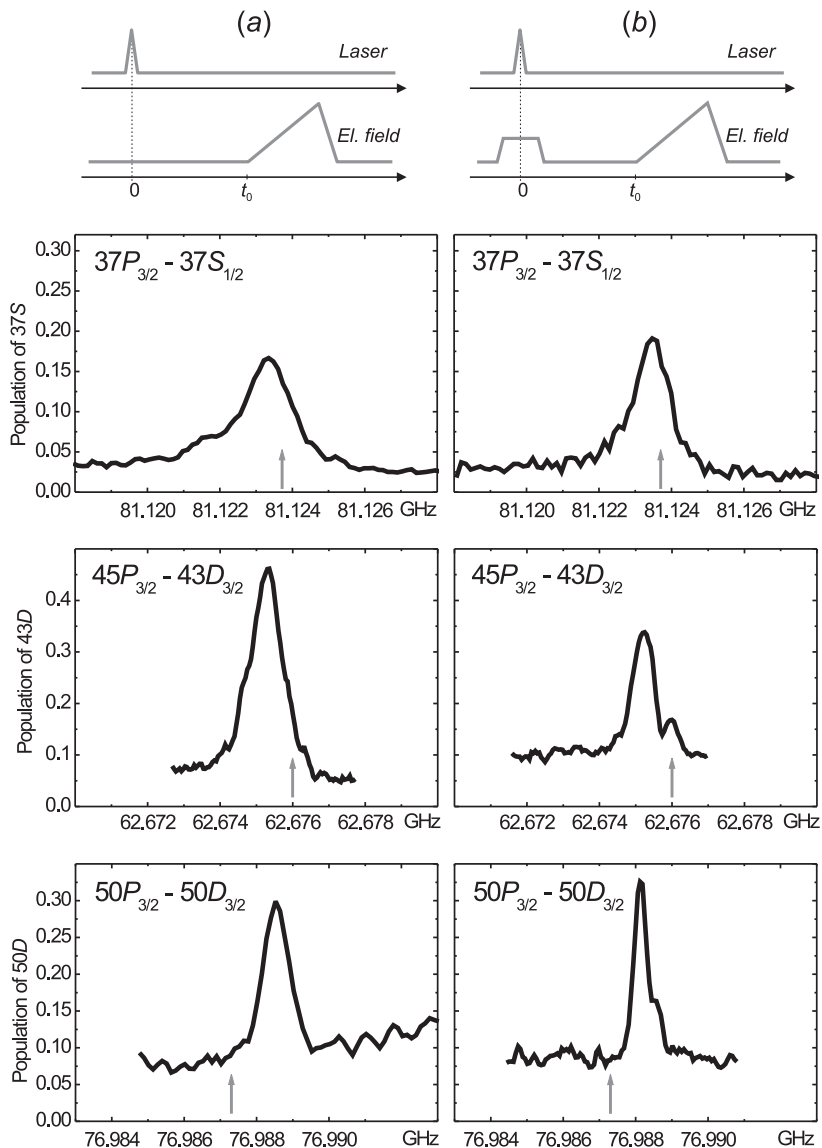}
\caption{\label{Fig6} Spectra of various microwave resonances between Rydberg states of Rb atoms: (\textit{a}) In the presence of cold photoions, which appeared at laser excitation of the initial \textit{nP} Rydberg state. The electric field of the photoions noticeably broadens the microwave resonances. (\textit{b}) Additional electric pulse (5 V/cm, 2~$\mu $s) at the moment of laser excitation rapidly extracts photoions, resulting in a narrowing of the microwave resonances. The arrows indicate the calculated positions of the resonances.}
\end{figure*}

The application of an additional extracting pulse for photoions allowed us to substantially reduce the width of microwave resonances [Fig.~6(b)] to 1.1 MHz for the transition $37P_{3/2} \to 37S_{1/2} $, to 0.64 MHz for the transition $45P_{3/2} \to 43D_{3/2} $, and to 0.38 MHz for $50P_{3/2} \to 50D_{3/2} $. In the case of the transition $37P_{3/2} \to 37S_{1/2} $ one should take into account that in the $^{85}$Rb isotope this transition originally has a hyperfine structure with a total width of about 0.44 MHz [9]. In addition, the power broadening by a microwave field also contributes to the total width of each resonance, because the Rabi frequency for the investigated resonance amounts to $\Omega /(2\pi )=$0.1$-$0.2~MHz. With regard to these sources of line broadening, the widths of microwave transitions in Fig.~6(b) are close to the least possible ones. Our experiments have shown that a necessary condition for this is a fast removal of photoions from the interaction region.

In spite of the fact the ultimate line width is reached in Fig.~6(b), the measured transition frequencies remain shifted with respect to the calculated values by $-$0.23~MHz for the transition $37P_{3/2} \to 37S_{1/2} $, by $-$0.73~MHz for the transition $45P_{3/2} \to 43D_{3/2} $, and by +0.81~MHz for the transition $50P_{3/2} \to 50D_{3/2} $, whereas the absolute accuracy of frequency measurements in our experiments is 0.1$-$0.2 MHz. This means that there is an uncompensated transverse electric field in the detection system. Applying a compensating voltage to the plates of the detection system, we can compensate only for the vertical component of the spurious electric field. The transverse field component, which arises due to the penetration of the field of the deflecting electrode or due to the contact potentials, remains uncompensated. This is indirectly confirmed by the fact that for the transitions $37P_{3/2} \to 37S_{1/2} $ and $45P_{3/2} \to 43D_{3/2} $ the shift is to lower frequencies, whereas for the transition $50P_{3/2} \to 50D_{3/2} $ the shift is to higher frequencies, in full agreement with the signs of the polarizability differences of these levels in a weak electric field. If our measurements had a systematic error, all the resonances would be shifted to the same side and by the same value. Moreover, along with the calculated values of the polarizabilities of Rydberg levels, the above shifts of the transition frequencies give the same value of the spurious transverse electric field, equal to $0.12\pm 0.2$ V/cm, which also indirectly confirms our interpretation of these shifts.

We can conclude that, to increase the accuracy of spectroscopic measurements on microwave transitions in cold Rydberg atoms, one should, in addition to removing photoions, carefully compensate for spurious electric fields in all coordinates. To this end, one should equip the detection system with additional electrodes that would allow one to vary the sign and the magnitude of the transverse compensating field in the interaction region. In [4], such compensation, combined with fast switching off the magnetic field of a MOT for the interaction time with microwave radiation, allowed the authors to observe resonances with a width of about 30 kHz.

In conclusion, we note that the method of microwave spectroscopy of Rydberg atoms was first applied to measuring weak electric fields in ultracold plasma of Rb atoms in [12]. In that study, the magnetic field of a MOT was not switched off; therefore, a one-photon microwave resonance had a width of 5 MHz, and the limit sensitivity of measuring the microscopic field of the plasma was 0.1~V/cm. To observe the shift of the microwave resonance, one needed about 10$^4$ Rb ions, because the size of the excitation volume was greater than 100 $\mu $m. The distinctive feature of our experiments is the use of a narrower resonance at the point of zero magnetic field in a MOT and thereby a higher sensitivity to weak electric fields (of about 10 mV/cm), the possibility of detecting microscopic fields form a small number (1$-$10) of photoions, and the higher spatial resolution (20$-$30 $\mu $m) due to the localization of the excitation volume in the geometry of tightly focused crossed laser beams. 

\section {CONCLUSIONS}

Based on the experimental data obtained, we can conclude that one of the fundamental sources of line broadening in ensembles of cold Rydberg atoms are photoions that are generated under the excitation of Rydberg states by broadband pulsed laser radiation. Since photoionization does not require resonance radiation, all the photons of pulsed lasers take part in photoionization, whereas only resonance photons take part in the excitation of Rydberg states; therefore, the photoionization probability turns out to be comparable with the probability of excitation of Rydberg atoms.

Photoionization gives rise to cold photoions that may stay in the excitation volume for a long time. Due to the Stark effect, the presence of cold photoions leads to the deviation of the frequency of atomic transitions during the interaction time, thus leading to asymmetric broadening of the F\"orster and microwave resonances. On the other hand, asymmetric broadening allows one nondestructively measure the mean field of ultracold plasma in a gas of cold Rydberg atoms. Moreover, our method for detecting charged particles makes it possible to measure the number of photoions generated in every laser pulse. The possibility of direct measurement of the mean electric field and the number of photoions may be of interest when studying various stages of formation of ultracold plasma in ensembles of Rydberg atoms.

The obtained data on the line shapes of microwave and F\"orster resonances in Rydberg atoms can also be used for developing various methods of narrowing the lines and increasing the measurement accuracy of transition frequencies. For example, a method to eliminate the unwanted generation of photoions consists in using a lower power narrow-band continuous-wave laser radiation to excite Rydberg states. In our case, a decrease in the line width of laser radiation from 10 GHz to 1 MHz would lead to a 10$^4$ time decrease in the generation probability of photoions, whereby the latter will hardly affect the observed spectra. This will allow one to analyze other possible sources of formation of ultracold plasma in dense ensembles of cold Rydberg atoms, such as the photoionization of Rydberg atoms by background thermal radiation [18], the Penning ionization under dipole-dipole interaction of Rydberg atoms [19], and the photoionization of cold atoms by laser radiation [20].

\begin{acknowledgments}
This work was supported by the Russian Foundation for Basic Research (Grant Nos. 10-02-00133, 09-02-90427, and 09-02-92428), by the programs of the Russian Academy of Sciences and of the Siberian Branch of the Russian Academy of Sciences, by the grants of the President of the Russian Federation (project Nos. MK-6386.2010.2 and MK-3727.2011.2), by the Dynasty Foundation, and by the project FP7-PEOPLE-2009-IRSES "COLIMA."
\end{acknowledgments}

\end{document}